\documentclass[12pt]{article}
\usepackage{amssymb}
\makeatletter
\@addtoreset{equation}{section}
\makeatother

\begin{document}
\begin{center}
{\Large \bf The Indeterministic Einstein Equation:\\[0.6cm]
Quantum Jumps, Spacetime Structure,
  and Dark Pseudomatter\\[1.5cm]}
{\bf Vladimir S.~MASHKEVICH}\footnote {E-mail:
Vladimir\_Mashkevich@qc.edu}  \\[1.4cm]
{\it
Physics Department\\
Queens College\\
The City University of New York\\
65-30 Kissena Boulevard\\
Flushing, New York 11367-1519}\\[1.4cm]
\vskip 1cm

{\large \bf Abstract}
\end{center}

Current physics is faced with the fundamental problem of unifying
quantum theory and general relativity, which would have resulted
in quantum gravity. The main effort to construct the latter has
been bent on quantizing spacetime structure, in particular metric.
Meanwhile, taking account of the indeterministic aspect of the
quantum description of matter, which manifests itself in quantum
jumps, essentially affects classical spacetime structure and the
Einstein equation. Quantum jumps give rise to a family of sets of
simultaneous events, which implies the existence of universal
cosmological time. In view of the jumps, the requirement for
metric and its time derivative to be continuous implies that the
Einstein equation should involve pseudomatter along with matter.
Pseudomatter  manifests itself only in gravitational effects,
being thereby an absolutely dark ``matter''.
\newpage

\section*{Introduction}

Current physics is faced with the fundamental problem of
constructing a unified physical theory. The problem boils down to
the unification of quantum theory and general relativity, which
would have resulted in quantum theory of gravitation, or quantum
gravity. The effort to construct quantum gravity has been bent,
for the most part, on quantizing spacetime structure, in
particular metric. Meanwhile, taking account of the
indeterministic aspect of the quantum treatment of matter, which
manifests itself in quantum jumps, essentially affects classical
spacetime structure and the Einstein equation. The reasoning is as
follows.

In the classical description of spacetime and quantum treatment of
matter, which is known as semiclassical gravity, the Einstein
equation reads $G_{\mu\nu}-\Lambda g_{\mu\nu}=8\pi\varkappa
T_{\mu\nu}$, $T_{\mu\nu}=(\Psi,\hat T_{\mu\nu}\Psi)$, where $G$ is
the Einstein tensor, $\Lambda$ is the cosmological constant, $g$
is metric, $\varkappa$ is the gravitational constant, $\hat T$ is
the energy-momentum tensor operator, and $\Psi$ is a state vector.
Quantum jumps of $\Psi$ have as a consequence jumps of
$T_{\mu\nu}$, which implies those of $\tilde G_{\mu\nu}\equiv
G_{\mu\nu}-\Lambda g_{\mu\nu}$. The components $G_{ij}\quad
(i,j=1,2,3)$ involve the second time derivatives $\ddot g_{ij}
\equiv g_{ij,00}$ of the metric components $g_{ij}$ [1,2].
Therefore the six equations $\tilde G_{ij}=8\pi\varkappa T_{ij}$
may remain unchanged: Jumps of $T_{ik}$ will result in those of
$\ddot g_{ij}$, which is physically appropriate. A completely
different type of situation occurs in the four equations $\tilde
G_{0\mu}=8\pi\varkappa T_{0\mu}$. The only time derivatives of
metric involved in the components $G_{0\mu}$ are $\dot g_{ij}$,
which should be continuous, let alone $g_{\mu\nu}$. Therefore the
four equations are violated by jumps of $T_{0\mu}$ and must be
extended. An apparent approach to the problem is to write $\tilde
G_{0\mu}=8\pi\varkappa(T_{0\mu}+ P_{0\mu})$ where $P_{0\mu}$ are
to compensate for the jumps of $T_{0\mu}$. We shall treat the
quantities $P_{0\mu}$ as the components of the pseudomatter
energy-momentum tensor.

There immediately arises the uniqueness problem: In order that
introducing $P_{0\mu}$ make sense, the separations of tensors into
$(0\mu)$ and $(ij)$ components should be unique. This implies the
existence of a distinguished universal time. It is the quantum
jumps that provide the existence of such a time. A quantum jump
gives rise to a set of simultaneous events---jumps of $\ddot g$.
These events allow for synchronizing clocks and thereby furnishing
universal time. It is natural to identify cosmological time with
quantum-jump universal time. Thus cosmological time is defined on
the level of fundamental physical laws---in contrast to the
phenomenological approach in classical cosmology[3,4].

Universal time gives rise to the product structure of spacetime
manifold, $M=T\times S$ where $T$ is cosmological time and $S$ is
cosmological space, and a particular synchronous reference
frame--- cosmological reference frame.

There remains the problem of pseudomatter dynamics, i.e., the time
dependence of the quantities $P_{0\mu}$. The latter is determined
by the dynamics of the tensor $T_{\mu\nu}$.

The foregoing results in the indeterministic Einstein equation:
$G_{\mu\nu}-\Lambda g_{\mu\nu}=S_{\mu\nu},\quad S_{\mu\nu}=
T_{\mu\nu}+P_{\mu\nu}$, $S$ stands for source, $T$ relates to
matter and $P$ to pseudomatter, respectively; in the cosmological
reference frame $P_{ij}=0$.

Pseudomatter manifests itself only in gravitational effects,
representing therefore an absolutely dark ''matter''.

The application of the indeterministic Einstein equation to the
Robertson-Walker spacetime results in that the source density is
$\rho_s=\rho_m+\rho_{ps}$ where $m$ relates to matter and $ps$ to
pseudomatter, respectively, in accordance with which the parameter
$\Omega=\rho/\rho_{cr}$ equals $\Omega_s=\Omega_m+\Omega_{ps}$.

\section{Quantum jumps and violation of the Einstein equation}

We adopt the classical description of spacetime and quantum
treatment of matter. The Einstein equation takes the form:
\begin{equation}
\label{(1.1)}
G-\Lambda g=8\pi\varkappa T
\end{equation}
\begin{equation}
\label{(1.2)}
T=(\Psi,\hat T\Psi)
\end{equation}
where $G$ is the Einstein tensor, $g$ is metric, $\Lambda$ is the
cosmological constant, $\varkappa$ is the gravitational constant,
$\hat T$ is the energy-momentum tensor operator, and $\Psi$ is a
state vector. An essential aspect of this treatment is quantum
indeterminism, which manifests itself in jumps of the state
vector:
\begin{equation}
\label{(1.3)}
\Psi_{\rm before\, jump}\equiv \Psi^<\to \Psi^>\equiv
\Psi_{\rm after\, jump}
\end{equation}
A jump of $\Psi$ results in that of $T$:
\begin{equation}
\label{(1.4)}
\Delta T=(\Psi^>,\hat T\Psi^>)-(\Psi^<,\hat T\Psi^<)
\end{equation}
under the assumption that $\hat T$ is continuous. Discontinuity of
$T$ causes a violation of the Einstein equation (1.1). Let us
consider the violation in detail. Write down equations (1.1),
(1.2), and (1.4) in components:
\begin{equation}
\label{(1.5)}
G_{\mu\nu}-\Lambda g_{\mu\nu}=8\pi\varkappa T_{\mu\nu},\quad
T_{\mu\nu}=(\Psi,\hat T_{\mu\nu}\Psi)
\end{equation}
\begin{equation}
\label{(1.6)}
\Delta T_{\mu\nu}=(\Psi^>,\hat T_{\mu\nu}\Psi^>)
-(\Psi^<,\hat T_{\mu\nu}\Psi^<)
\end{equation}
Let a jump of $\Psi$ occur at $x^0=t=t_{\rm jump}$. We put
\begin{equation}
\label{(1.7)} \Psi^<=\Psi(t_{\rm jump}),\quad\Psi^>=\Psi(t_{\rm
jump}+0)
\end{equation}
so that
\begin{equation}
\label{(1.8)} \Delta T_{\mu\nu}(t_{\rm jump},\vec
x)=T_{\mu\nu}(t_{\rm jump}+0,\vec x)- T_{\mu\nu}(t_{\rm jump},\vec
x),\quad\vec x=(x^i),\:i=1,2,3
\end{equation}
The components $G_{ij}\;(i,j=1,2,3)$ of the Einstein tensor
involve the second time derivatives
\begin{equation}
\label{(1.9)}
\ddot g_{ij}\equiv g_{ij,00}
\end{equation}
of the metric tensor components $g_{ij}$ [1,2]. This makes it
possible to retain the six equations
\begin{equation}
\label{(1.10)}
G_{ij}-\Lambda g_{ij}=8\pi\varkappa T_{ij}
\end{equation}
unchanged. Jumps of the $T_{ij}$ will result in those of the
$\ddot g_{ij}$, which is quite conceivable from the physical point
of view: A jump of force results in a jump of acceleration. As to
the four equations
\begin{equation}
\label{(1.11)}
G_{0\mu}-\Lambda g_{0\mu}=8\pi\varkappa T_{0\mu}
\end{equation}
the situation is completely different. The components $G_{0\mu}$
of the Einstein tensor involve no second time derivatives of
metric tensor components; the only time derivatives of metric
involved in $G_{0\mu}$ are $\dot g_{ij}$. The latter should be
continuous, not to mention $g_{\mu\nu}$. Thus the violation of the
four equations (1.11) is intolerable and they must be extended.

\section{Introducing pseudomatter}

An apparent approach to the improvement of equations (1.11) is to
write
\begin{equation}
\label{2.1}
G_{0\mu}-\Lambda g_{0\mu}=8\pi\varkappa(T_{0\mu}+P_{0\mu}),\quad
\mu=0,1,2,3
\end{equation}
in which the terms $P_{0\mu}$ are to compensate for the jumps of
$T_{0\mu}$. Dynamical equations are (1.10), whereas (1.11) play
the role of integrals of motion, so that introducing the terms
$P_{0\mu}$ into (2.1) is quite justified. We treat the quantities
$P_{0\mu}$ as the components of the energy-momentum tensor
relating to pseudomatter.

\section{The problem of uniqueness of pseudomatter}

There immediately arises the uniqueness problem.
In order that the above
procedure of introducing the terms $P_{0\mu}$ make sense, the
separation of the tensors in question into $(0\mu)$ and $(ij)$
parts
needs to be unique. This, in its turn, implies the existence of a
distinguished $x^0$ coordinate, or a universal time. Cosmological
time, i.e., universal time of the standard model of cosmology
cannot be taken for this purpose, for it is introduced
phenomenologically rather than on the basis of fundamental
physical laws [3,4].

\section{Quantum-jump universal time}

It is quantum jumps that appear again, this time to provide a
universal time. A quantum jump of the state vector gives rise to a
set of events---jumps of $\ddot g$. These events are, by
definition, simultaneous, which allows for synchronizing clocks
and thereby furnishing the universal time. It is natural to
identify phenomenological cosmological time with the quantum-jump
universal time. Now cosmological time is defined on the level of
fundamental physical laws.

In special relativity the concept of simultaneity in connection
with quantum jumps makes no operationalistic sense. Taking
gravity into account endows the concept with an operationalistic
content.

\section{Spacetime structure}

Universal time gives rise to a family of sets of simultaneous
events, thereby endowing spacetime with a fiber structure. The
metric compatible with this structure, i.e., admitting
synchronization of clocks, is of the form [2]
\begin{equation}
\label{5.1}
ds^2=g_{00}(dx^0)^2+g_{ij}dx^idx^j
\end{equation}
or, with
\begin{equation}
\label{5.2}
dt=g_{00}dx^0,\quad t=t(x^0,\vec x)=\int g_{00}(x^0,\vec x)dx^0
\end{equation}
\begin{equation}
\label{5.3}
ds^2=dt^2+g_{ij}dx^idx^j,\quad g_{ij}=g_{ij}(t,\vec x)
\end{equation}
which relates to a synchronous reference frame. The latter, in its
turn, implies the product spacetime manifold:
\begin{equation}
\label{5.4}
M=M^4=T\times S,\quad M\ni p=(t,s),
\quad t\in T,\quad -\infty\le a\le t\le b
\le \infty,\quad s\in S
\end{equation}
The one-dimensional manifold $T$ is universal {cosmological} time,
the three-dimensional manifold $S$ is cosmological space. By
(5.4), the tangent space $M_p$ at a point $p\in M$ is
\begin{equation}
\label{5.5}
M_p=T_t\oplus S_s,\quad p=(t,s)
\end{equation}
and, in view of (5.3),
\begin{equation}
\label{5.6}
T_t\perp S_s
\end{equation}

Thus, quantum jumps give rise to the product spacetime (5.4) and a
particular synchronous reference frame
\begin{equation}
\label{5.7}
p=(t,s)\leftrightarrow (t,\vec x)
\end{equation}
The latter may be called cosmological reference frame and
considered as a canonical synchronous reference frame.

In the coordinate-free representation, the metric (5.3) reads
\begin{equation}
\label{5.8}
g=dt\otimes dt-h_t,\quad h_t\leftrightarrow -g_{ij}(t,\vec x)
dx^idx^j
\end{equation}
in which $h_t$ is a Riemannian metric tensor on $S$ depending
on $t$.

\section{Pseudomatter dynamics}

In order for equations (2.1) not to be merely a definition of the
quantities $P_{0\mu}$:
\begin{equation}
\label{6.1}
P_{0\mu}\equiv \frac1{8\pi\varkappa}(G_{0\mu}-\Lambda g_{0\mu})
-T_{0\mu}
\end{equation}
pseudomatter dynamics, i.e., time evolution of these quantities
should be determined independently of the dynamics for the
right-hand side of (6.1).

We have
\begin{equation}
\label{6.2} G_{\mu\nu}-\Lambda g_{\mu\nu}=8\pi\varkappa
S_{\mu\nu},\quad S_{\mu\nu}=S_{\nu\mu}
\end{equation}
where $S$ stands for source (and has nothing to do with the $S$ in
(5.4)). In the canonical synchronous reference frame, which will
be used henceforth,
\begin{equation}
\label{6.3}
S_{ij}=T_{ij},\quad S_{0\mu}=T_{0\mu}+P_{0\mu},\quad P_{ij}=0
\end{equation}
and
\begin{equation}
\label{6.4}
g_{0i}=0,\quad g_{00}=1
\end{equation}
 From (6.2) follows
\begin{equation}
\label{6.5}
S_\mu{}^\nu{}_{;\nu}=0
\end{equation}
For any symmetric tensor $Y^{\mu\nu}=Y^{\nu\mu}$
\begin{equation}
\label{6.6}
Y_\mu{}^\nu{}_{;\nu}\sqrt{-g}=(Y_\mu{}^\nu\sqrt{-g})_{,\nu}-
(1/2)g_{\alpha\beta,\mu}Y^{\alpha\beta}\sqrt{-g},\quad g={\rm
det}(g_{\mu\nu})
\end{equation}
holds [5]. We find from (6.5), (6.6), (6.4), and (5.3)
\begin{equation}
\label{6.7}
(S_\mu{}^0\sqrt{-g})_{,0}=(1/2)g_{kl,\mu}T^{kl}\sqrt{-g}-
(S_\mu{}^j\sqrt{-g})_{,j}
\end{equation}
whence
\begin{equation}
\label{6.8}
S_\mu{}^0\sqrt{-g}=\int_{t_0}^t dt'\{(1/2)g_{kl,\mu}T^{kl}
\sqrt{-g}-(S_{\mu}{}^j\sqrt{-g})_{,j}\}+f_\mu(\vec x)
\end{equation}
Thus
\begin{equation}
\label{6.9}
S_{0\mu}(t,\vec x)=\frac1{\sqrt{-g(t,\vec x)}}[F_\mu(t,\vec x)+
f_\mu(\vec x)]
\end{equation}
or, for brevity,
\begin{equation}
\label{6.10}
S_{0\mu}=\frac1{\sqrt{-g}}[F_\mu+f_\mu]
\end{equation}
where
\begin{equation}
\label{6.11}
F_\mu=\int_{t_0}^t dt'\{(1/2)g_{kl,\mu}T^{kl}\sqrt{-g}-
(S_\mu{}^j\sqrt{-g})_{,j}\}
\end{equation}
so that, in view of (6.3), (6.4),
\begin{equation}
\label{6.12}
F_i=\int_{t_0}^t dt'\{(1/2)g_{kl,i}T^{kl}\sqrt{-g}-
(T_i{}^j\sqrt{-g})_{,j}\}
\end{equation}
\begin{equation}
\label{6.13}
\begin{array}{l}
{\displaystyle
F_0=\int_{t_0}^t dt'\{(1/2)g_{kl,0}T^{kl}\sqrt{-g}-
(g^{ji}S_{0i}\sqrt{-g})_{,j}\}}\\
\qquad{\displaystyle
=\int_{t_0}^t dt'\{(1/2)g_{kl,0}T^{kl}\sqrt{-g}-
(g^{ji}[F_i+f_i])_{,j}\}}
\end{array}
\end{equation}
As to initial conditions, we have
\begin{equation}
\label{6.14}
F_\mu(t_0)=0
\end{equation}
so that
\begin{equation}
\label{6.15}
f_\mu=f_\mu(\vec x)=\frac1{8\pi\varkappa}[\sqrt{-g}(G_{0\mu}-
\Lambda g_{0\mu})](t_0,\vec x)
\end{equation}

In view of (6.3), the dynamics of $P_{0\mu}$ reduces to that of
$S_{0\mu}$. The latter is described by equations (6.10), (6.12),
(6.13), and (6.15).

Let us explicitly introduce jumps $\Delta T_{0\mu}$ into
consideration. It follows from (6.6) that
\begin{equation}
\label{6.16}
(1/2)g_{kl,\mu}T^{kl}\sqrt{-g}-(T_{\mu}{}^j\sqrt{-g})_{,j}=
(T_\mu{}^0\sqrt{-g})_{,0}-T_\mu{}^\nu{}_{;\nu}\sqrt{-g}
\end{equation}
Now we find from (6.12)
\begin{equation}
\label{6.17}
F_i=\sqrt{-g}\,T_{0i}-[\sqrt{-g}\,T_{0i}](t_0)-\sum_{t_0\le t_k<t}
[\sqrt{-g}\,\Delta T_{0i}](t_k)-\int_{t_0}^t dt'
T_i{}^\nu{}_{;\nu}\sqrt{-g}
\end{equation}
and from (6.13), (6.10), (6.17), and (6.15)
\begin{equation}
\label{6.18}
\begin{array}{l}
{\displaystyle
F_0=\sqrt{-g}\,T_{00}-[\sqrt{-g}\,T_{00}](t_0)-
\sum_{t_0\le t_k<t}[\sqrt{-g}\,\Delta T_{00}](t_k)-
\int_{t_0}^t dt'T_0{}^\nu{}_{;\nu}\sqrt{-g}}\\
{\displaystyle
-\int_{t_0}^t dt'\left(g^{ji}
\left\{\left[\sqrt{-g}\left(\frac1{8\pi
\varkappa}G_{0i}-T_{0i}\right)\right](t_0)-
\sum_{t_0\le t_k<t}[\sqrt{-g}\,\Delta T_{0i}](t_k)
-\int_{t_0}^{t'}dt''T_i{}^\nu{}_{;\nu}\sqrt{-g}
\right\}\right)_{,j}}
\end{array}
\end{equation}
Thus we obtain from (6.3), (6.10), (6.17), (6.18), and (6.15)
\begin{equation}
\label{6.19}
\begin{array}{l}
{\displaystyle
P_{0\mu}=\frac1{8\pi\varkappa\sqrt{-g}}
[\sqrt{-g}\,(G_{0\mu}-\Lambda g_{0\mu}-8\pi\varkappa
  T_{0\mu})](t_0)}\\
\qquad {\displaystyle -\frac1{\sqrt{-g}} \left\{\int_{t_0}^t
dt'\sqrt{-g}\,T_\mu{}^\nu{}_{;\nu} +\sum_{t_0\le
t_k<t}[\sqrt{-g}\,\Delta T_{0\mu}](t_k)\right\}}\\ \qquad
{\displaystyle -\frac{g_{0\mu}}{8\pi\varkappa\sqrt{-g}}
\int_{t_0}^t dt'\{g^{ji}[\sqrt{-g}\, (G_{0i}-8\pi\varkappa
T_{0i})](t_0)\}_{,j}}\\ \qquad {\displaystyle
+\frac{g_{0\mu}}{\sqrt{-g}}\int_{t_0}^t dt'\left\{g^{ji}
\left(\int_{t_0}^{t'}dt''\sqrt{-g}\,T_i{}^\nu{}_{;\nu}
+\sum_{t_0\le t_k<t} [\sqrt{-g}\,\Delta
T_{0i}](t_k)\right)\right\}_{,j}}
\end{array}
\end{equation}
This explicitly describes pseudomatter dynamics.

Note that the components $P_{ij}$ cannot be introduced, for there
are only four pseudomatter dynamics equations (6.5) and they
determine  the components $P_{0\mu}$.

\section{The indeterministic Einstein equation}

We have obtained the indeterministic Einstein equation
\begin{equation}
\label{7.1}
G-\Lambda g=8\pi\varkappa S
\end{equation}
In canonical synchronous reference frame it reads:
\begin{equation}
\label{7.2}
G_{ij}-\Lambda g_{ij}=8\pi\varkappa T_{ij}
\end{equation}
and (2.1) with $P_{0\mu}$ (6.19), i.e.,
\begin{equation}
\label{7.3}
\begin{array}{l}
G_{0\mu}-\Lambda g_{0\mu}-8\pi\varkappa T_{0\mu}\\
\qquad
{\displaystyle
+\frac{8\pi\varkappa}{\sqrt{-g}}\left\{\int_{t_0}^t dt'
\sqrt{-g}\,T_\mu{}^\nu{}_{;\nu}+\sum_{t_0\le t_k<t}
[\sqrt{-g}\,\Delta T_{0\mu}](t_k)\right\}}\\
\qquad
{\displaystyle
+\frac{g_{0\mu}}{\sqrt{-g}}\int_{t_0}^t dt'\{g^{ji}[\sqrt{-g}
(G_{0i}-8\pi\varkappa T_{0i})](t_0)\}_{,j}}\\
\qquad
{\displaystyle
-\frac{8\pi\varkappa g_{0\mu}}{\sqrt{-g}}
\int_{t_0}^t dt'\left\{g^{ji}
\left(\int_{t_0}^{t'}dt''\sqrt{-g}\,T_i{}^\nu{}_{;\nu}
+\sum_{t_0\le t_k<t}[\sqrt{-g}\,\Delta T_{0i}](t_k)\right)
\right\}_{,j}}\\
\qquad
{\displaystyle
=\frac1{\sqrt{-g}}[\sqrt{-g}\,(G_{0\mu}-\Lambda g_{0\mu}-
8\pi\varkappa T_{0\mu})](t_0)}
\end{array}
\end{equation}
Six equations (7.2) are dynamical ones whereas four equations
(7.3) play the role of integrals of motion.

A complete system of dynamical equations should include an
equation for the state vector $\Psi$.

Initial conditions at $t=t_0$ are given by
\begin{equation}
\label{7.4}
g_{ij}(t_0,\vec x),\quad \dot g_{ij}(t_0,\vec x),\quad\Psi_{t_0}
\end{equation}

The classical Einstein equations (1.11) are obtained by putting
\begin{equation}
\label{7.5}
\Delta T_{0\mu}=0, \quad T_\mu{}^\nu{}_{;\nu}=0,\quad
[G_{0\mu}-\Lambda g_{0\mu}-8\pi\varkappa T_{0\mu}](t_0)=0
\end{equation}
in (7.3). In order that the complete system (1.10), (1.11) be
fulfilled in all reference frames, the condition
\begin{equation}
\label{7.6}
\Delta T_{ij}=0
\end{equation}
must hold as well.

\section{Pseudomatter as an absolutely dark ``matter''}

There exists the well known problem of dark matter [6,7,8]. There
is no interaction between pseudomatter and matter, except that
they interact through gravity. Thus pseudomatter manifests itself
only in gravitational effects and may therefore be regarded as an
absolutely dark ``matter''.

\section{Pseudomatter in the Robertson-Walker spacetime}

Let us consider $P_{0\mu}$ (6.19) in the case of the
Robertson-Walker spacetime. In this case
\begin{equation}
\label{9.1}
G_{0i}=0,\quad T_{0i}=0
\end{equation}
so that (7.3) for $\mu=i$ results in
\begin{equation}
\label{9.2}
T_i{}^\nu{}_{;\nu}=0
\end{equation}
Thus
\begin{equation}
\label{9.3}
P_{0i}=0
\end{equation}
\begin{equation}
\label{9.4}
\begin{array}{l}
{\displaystyle P_{00}=\frac1{8\pi\varkappa\sqrt{-g}}
[\sqrt{-g}(G_{00}-\Lambda-8\pi\varkappa T_{00})](t_0)}\\
{\displaystyle -\frac1{\sqrt{-g}}\left\{\int_{t_0}^t
dt'\sqrt{-g}\, T_0{}^\nu{}_{;\nu} +\sum_{t_0\le
t_k<t}[\sqrt{-g}\,\Delta T_{00}](t_k)\right\}}
\end{array}
\end{equation}
Next we assume that matter energy $E$ is continuous:
\begin{equation}
\label{9.5}
\Delta E(t_k)=0
\end{equation}
then
\begin{equation}
\label{9.6}
\Delta T_{00}(t_k)=0
\end{equation}
In view of (9.2) we put
\begin{equation}
\label{9.7}
T_0{}^\nu{}_{;\nu}=0
\end{equation}
so that
\begin{equation}
\label{9.8}
T_\mu{}^\nu{}_{;\nu}=0
\end{equation}
Now (9.4) reduces to
\begin{equation}
\label{9.9}
P_{00}=\frac1{8\pi\varkappa}\frac{\sqrt{-g(t_0)}}{\sqrt{-g}}\,
[G_{00}-\Lambda-8\pi\varkappa T_{00}](t_0)
\end{equation}
We have
\begin{equation}
\label{9.10}
\frac{\sqrt{-g(t_0)}}{\sqrt{-g}}=\frac{R^3(t_0)}{R^3(t)}
\end{equation}
where $R$ is the radius of the universe.

Thus the pseudomatter density
\begin{equation}
\label{9.11}
\rho_{ps}=\rho_{ps}(t)=P_{00}=\frac
{[R^3(G_{00}-\lambda-8\pi\varkappa\rho_m)](t_0)}{8\pi\varkappa}
\frac1{R^3(t)}\propto\frac1{R^3(t)}
\end{equation}
where
\begin{equation}
\label{9.12}
\rho_m=T_{00}
\end{equation}
is the matter density. The source density is
\begin{equation}
\label{9.13}
\rho_s=\rho_m+\rho_{ps}
\end{equation}
in accordance with which, the parameter $\Omega=\rho/\rho_{cr}$ is
equal to
\begin{equation}
\label{9.14}
\Omega_s=\Omega_m+\Omega_{ps}
\end{equation}
These results match those of [9].

\section*{Acknowledgments}

I would like to thank Alex A. Lisyansky for support and
Stefan V. Mashkevich for helpful discussion.

\end{document}